%Paper: gr-qc/9208003
%From: tsg@iucaa.ernet.in (Ghosh)
%Date: Tue, 11 Aug 92 22:30:09 GMT

%\magnification=\magstep1
\hsize 6.0 true in
\vsize 9.0 true in
\voffset =  -.2 true in
\font\tentworm=cmr10 scaled \magstep2
\font\tentwobf=cmbx10 scaled \magstep2

\font\tenonerm=cmr10 scaled \magstep1
\font\tenonebf=cmbx10 scaled \magstep1

\font\eightrm=cmr8
\font\eightit=cmti8
\font\eightbf=cmbx8
\font\eightsl=cmsl8
\font\sevensy=cmsy7
\font\sevenm=cmmi7

\font\twelverm=cmr12
\font\twelvebf=cmbx12
\def\subsection #1\par{\noindent {\bf #1} \noindent \rm}

\def\mid {\let\rm=\tenonerm \let\bf=\tenonebf \rm \bf}

\def\para{\par \vskip 12 pt}

\def\head{\let\rm=\tentworm \let\bf=\tentwobf \rm \bf}

\def\heading #1 #2\par{\centerline {\head #1} \smallskip
 \centerline {\head #2} \vskip .15 pt \rm}

\def\eight{\let\rm=\eightrm \let\it=\eightit \let\bf=\eightbf
\let\sl=\eightsl \let\sy=\sevensy \let\m=\sevenm \rm}

\def\foots{\noindent \eight \baselineskip=10 true pt \noindent \rm}
\def\sexion{\let\rm=\twelverm \let\bf=\twelvebf \rm \bf}

\def\section #1 #2\par{\vskip 20 pt \noindent {\mid #1} \enspace {\mid #2}
  \para \noindent \rm}

\def\ssection #1 #2\par{\noindent {\mid #1} \enspace {\mid #2}
  \para \noindent \rm}

\def\abstract#1\par{\para \foots {\bf Abstract: \enspace}#1 \para}

\def\author#1\par{\centerline {#1} \vskip 0.1 true in \rm}

\def\abstract#1\par{\noindent {\bf Abstract: }#1 \vskip 0.5 true in \rm}

\def\midsection #1\par{\noindent {\sexion #1} \noindent \rm}

\def\sqr#1#2{{\vcenter{\vbox{\hrule height#2pt
 \hbox {\vrule width#2pt height#1pt \kern#1pt
  \vrule width#2pt}
  \hrule height#2pt}}}}

\baselineskip=24pt
\pretolerance=10000

%%%%%%%%%%%%%%%%%%%%%%%%%% GENERAL MACROS USED %%%%%%%%%%%%%%%%%%%%%%%%%%%

\def\I{\item}

\def\n{\noindent}
\def\s{\smallskip}
\def\m{\medskip}

%%%%%%%%%%%%%%%%%%%%%%%%%%%%%%%%%%%%%%%%%%%%%%%%%%%%%%%%%%%%%%%%%%%%%%%%%%

%%%%%%%%%%%%%%%%%%% MATHEMATICAL MACROS USED %%%%%%%%%%%%%%%%%%%%%%%%%%%%%
\def\ra { \rangle}
\def \la { \langle }
\def\mn{{\mu\nu}}
\def\m{\mu}

\def\e{\epsilon}

\def\d{\delta}
\def\Ora{\la\Omega \ra}
\def\oo{\Omega_o (t) }

\def\p{ \phi ({\bf x}, t) }
\def\t{\theta}
\def\T{\tau}
\def\pl{\partial}

\def\D{\Delta}

\def\O{\Omega}
\def\G{\Gamma}

\def\boxit#1{\vbox {\hrule \hbox {\vrule \kern3pt
\vbox{\kern3pt#1\kern3pt}\kern3pt\vrule}\hrule}}
\def\boxit#1{\vbox {\hrule \hbox {\vrule \kern3pt
\vbox{\kern3pt#1\kern3pt}\kern3pt\vrule}\hrule}}

\def \dal {\boxit {\phantom ,}}
%%%%%%%%%%%%%%%%%%%%%%%%% titlepage %%%%%%%%%%%%%%%%%%%%%%%%%%%%%%%%%%%%%%%

\line{\hfill \foots{IUCAA Preprint}}
\line{\hfill \foots{June 1992}}
\vskip .35 true in
\medskip

\heading { Generation of Seed Perturbations}

\heading {from Quantum Cosmology}

\smallskip

\centerline{by}
\smallskip
\centerline{Tarun Souradeep$^*$}
\medskip
\centerline{Inter-University Centre for Astronomy and Astrophysics}
\centerline{Post Bag 4, Ganeshkhind, Pune 411 007}
\centerline{INDIA}
\bigskip
\bigskip
\bigskip
\bigskip
\bigskip
\bigskip
\bigskip
\bigskip
\bigskip

\vfill
\settabs 10\columns
\+&&&$^*$ email~:&~~~~tarun@iucaa.ernet.in \cr

\vfill\eject
%%%%%%%%%%%%%%%%%%%%%%%%%%% abstract %%%%%%%%%%%%%%%%%%%%%%%%%%
\bigskip
\bigskip
\bigskip
\bigskip
\bigskip
\heading {Abstract}

\noindent The origin of seed perturbations in the Universe is studied
 within the framework of a specific minisuperspace model.
It is shown that the `creation' of the Universe as a result of a quantum
transition from a flat empty spacetime would lead to a flat  FLRW
(Friedmann Lema\^\i tre Robertson-Walker)
Universe with weak inhomogeneous  perturbations at large wavelengths.
 The power spectrum of these
perturbations is found to be scale invariant at horizon crossing
(i.e., the Harrison-Zeldovich spectrum).
It is also recognised  that the seed perturbations generated in our
 model would  be generically of the isocurvature kind.

\bigskip

\noindent {\bf subject headings: }~quantum cosmology; structure
formation; seed perturbations.

\vfill\eject
%%%%%%%%%%%%%%%%%%%%%%%%%%%%%%%%%%%%%%%%%%%%%%%%%%%%%%%%%%%%%%%%%%%%%%%%%
\section { 1. Introduction}

Classical models of cosmology till the turn of the last decade were
 unable to resolve
three fundamental problems, commonly known as  i) the singularity,
ii) the horizon problem (~ Misner 1969~),
and iii) the flatness problem (~Dicke and Peebles 1979~).

\n An exciting development in the early 80's was the emergence of
 the concept of inflation as a
promising remedy for the above problems (~Guth 1981~). Inflationary
 scenarios chiefly resolve
the horizon and the flatness problems though a few do address the
 singularity problem
(~Starobinsky 1980; Linde 1982~).
The problem of the origin of structure in the universe, which ought
to have been included in the above
list,  was relegated to an initial value problem until inflationary
 scenarios
 showed promises of successfully addressing it. It was then realised
 very soon that
all `natural models' of inflation produce density perturbations with
the right spectrum but with an
unacceptably large amplitude (~Hawking 1982; Starobinsky 1982; Guth
and Pi 1982; Bardeen 1987~). Since then, cosmology
 has seen the rise and fall of numerous inflationary scenarios with
various forms of fine-tuning to constrain
the amplitude of the perturbations (~Brandenberger 1985; Olive 1990;
 Narlikar and Padmanabhan 1991~). Even granted
the fine-tuned parameters needed to resolve the problem of density
 perturbations, the claim of inflation of generically producing  the
 flat ($ \rho_o / \rho_c \approx 1 $) FLRW
(Freidmann Lema\^\i tre  Robertson Walker)
universe  has also been questioned by Ellis (1988; also see Madsen
 and Ellis 1988~). These authors have pointed out that
$\rho_o / \rho_c \approx 1 $ is not a generic outcome of inflation
 and, in fact, has to be fine-tuned
 to its present value. Similarily, the horizon problem is also not
 permanently eliminated but remedied only for
the present epoch  pushing it to an epoch far in the future
 (~Padmanabhan and Seshadri 1988; Ellis and Stoeger 1988~).
Thus the strong points in favour of the inflationary
model which made it so attractive in the beginning now do not seem
 so strong. In particular, alternative scenarios
for the very early universe can certainly be tried.

\n An attempt to resolve the above important problems by taking
recourse to a
simplistic but working model of quantum cosmology was made by Narlikar
 and Padmanabhan
(1983).  This approach to quantum gravity  (~Narlikar 1979, 1981;
Padmanabhan and Narlikar 1982; Padmanabhan  1982~) is
based on a path integral formulation of the
quantum  version of classical geometrodynamics, where only the
 conformal degree
 of freedom  is quantised  leaving  the other degrees (extrinsic
curvature)
 frozen. Although this restriction may oversimplify the problem of
quantizing gravity, it has certain advantages.
First, an exact nonperturbative solution of the quantum geometrodynamic
 evolution is possible. Secondly, restriction
to conformal degrees of freedom only implies a quantum theory of
gravity where the  quantum fluctuations
preserve the causal structure of spacetime. The conformal quantisation
leads to many interesting results, e.g.,
cosmological spacetimes with singularity \footnote {$^\dagger$}
{ To be precise, we are refering to solutions with curvature
singularities,  i.e., points where the curvature invariants become
 unbounded.} (and the consequent horizon problem)
 appear to be a set of zero measure in the solution space and that
the flat FLRW universe happens to be
the most favoured conformally flat spacetime arising out of a quantum
transition from the Minkowskian spacetime.

\n These successes prompt us to take up a problem at the next level
of sophistication, viz. the origin of fluctuations
against a homogeneous background. In this paper we extend the above
approach to study the possible
generation of primordial density fluctuations in a universe created
 by quantum transition from a
 flat empty spacetime. To this end we briefly review the Narlikar -Padmanabhan
approach to quantum cosmology
in $\S$2  highlighting the resolution of the flatness problem that
 quantum conformal
transitions of the universe from a quantum state peaked around the
flat Minkowski
 spacetime  (~$\Ora_i \equiv 1   $~) would, with maximum probability,
 end up in a state peaked
around a flat FLRW universe (~$\Ora_f \equiv \oo $~). Then in $\S$3 we
show that fluctuations around a background $\oo$
in the form of inhomogeneous modes, $\p $, can be perturbatively
introduced into  the  mean conformal factor $\Ora$ with marginally
diminished probability and obtain the power spectrum of the
fluctuations $\p $. It is also demonstrated that  conformal fluctuations
 always imply  isocurvature perturbations. The translation of  the
 conformal fluctuation $\p$ to density  perturbations $\d M / M $ once
the hot radiation  dominated matter loses its conformal invariance, is
studied in $\S$4.

\n In $\S$5, we evolve the fluctuations $\p$ in a universe with some
form of coupling of  matter to
 the conformal degree of gravity in the matter Lagrangian. $\oo$ is
 obtained through the evolution equation for $\p$.
However, we outline in $\S $6, an alternative prescription for evolving
 the perturbations wherein the form of the homogeneous background
conformal factor $\oo$ is independently
fixed  which then determines the evolution of $\p$.

\n We end the paper with a discussion ($\S$7) comparing the results
 in the present `standard'
and `non-standard' scenarios of generation of primordial density
fluctuations.

%%%%%%%%%%%%%%%%%%%%%%%% Section 2
% %%%%%%%%%%%%%%%%%%%%%%%%%%%%%%%%%%%%%%%%%%%%%%%%%%%%%%%%%

\section {2. Quantum conformal fluctuations in Cosmology}

At a classical level, general relativity can be formulated as a
 dynamical theory by considering a
$3+1$ York decomposition (~York 1972~)  of the spacetime manifold $\cal M$ into
3-hypersurfaces $\Sigma (t) $
evolving along a timelike curve, parameterised by  a time $t$,
between the boundaries $\Sigma_i$
 and $\Sigma_f$ --- the specified initial and final hypersurfaces
at times $t_i$ and $t_f$.
The 3-geometry on $\Sigma $ at a given time $t$ is specified by the
scale factor $\O $ and
the extrinsic curvature $K_{ab}$. The classical solution to general
relativity is the trajectory
$\G_{cl} (t) $ in the superspace ${\cal G}$  of 3-geometries which
extremises the action

$$ J = {1 \over 16 \pi } \int_{\cal V} R~  \sqrt {-g}~ d^4x  ~+ J_m
\eqno (2.1) $$

\n (where  $J_m$ is the action for matter fields) over the  4-volume
 ${\cal V}$ between
$\Sigma_i$ and $\Sigma_f$ (~Isenberg and Wheeler 1979~).

\n At the quantum level, the above picture translates to calculating
 the probability amplitude
$K [ {\cal G}_i; {\cal G}_f]$ for transition from  a 3-geometry
 ${\cal G}_i$ on $\Sigma_i (t_i) $ to
${\cal G}_f$ on  $\Sigma_f (t_f) $. In exact analogy to the path
integral formulation of quantum mechanics (~Feynman and Hibbs 1965~),
 the
transition amplitude  $K [ {\cal G}_i; {\cal G}_f]$ can be  formally
 expressed as a sum over
all trajectories $\G (t)$ in the superspace ${\cal G}$ joining
${\cal G}_i$ and ${\cal G}_f$ as

$$ K [ {\cal G}_i; {\cal G}_f] = \sum_\G  {\rm exp }
 \bigg[ i {J[\G] \over \hbar} \bigg]~.
	\eqno(2.2)  $$

\n The evaluation of $K [ {\cal G}_i; {\cal G}_f]$, which contains
the complete
 essence of quantum gravity, is however beset with conceptual and
 technical difficulties.
The  expression in (2.2) gets considerably simplified and well
defined if one demands the
preservation of the  causal structure of spacetime for all paths $\G$.
 In this case, the paths $\G$ which are
summed over are such that the 4-geometries  $g_\mn (\G)$ along them
 are all conformally related to the classical
4-geometry along $\G_{cl}$ i.e., the metrics allowed are

$$ g_\mn (\G) =  \O^2 ({\bf x} , t)~ \tilde g_\mn(\G_{cl}) \eqno (2.3) $$

\n where $\O ({\bf x},t)$ is  a $C^2$ function of the spacetime
coordinates. This conformal degree of freedom
corresponds to the volume of the 3-hypersurface and has a special
status because it contributes a negative term
to the kinetic energy in the Wheeler-DeWitt  (W-D) equation.
 In line with Wheeler's philosophy that ``the 3-geometry is
the carrier of information about time " (~Kuch$\check{\rm a}$r 1971;
York 1971; Misner et al 1973~), the conformal factor
$\O ({\bf x} , t)$  can play the role of `time' to describe evolution
 in the superspace ${\cal G}$ through the
 `time-less' W-D equation. In fact, the conformal factor does appear
 to play the role of time
in the semi-classical regime of the W-D equation (~Padmanabhan 1989~).
 The conformal degree of freedom is sometimes
dismissed as `unphysical', a point of view with which we disagree.
In $\S$7, we shall deal with this particular issue.

\n The expression for $K [ {\cal G}_i; {\cal G}_f]$ in (2.2) thus
 reduces to

$$ K [ \O_i; \O_f] = \int {\cal D} \O ~ {\rm exp}~ \bigg[ {i \over
 12~l_p^2}	\int_{\cal V} (~\tilde R \O^2 - 6 \O^\mu \O_\mu )
 \sqrt {- \tilde g}~d^4x\bigg] ~.\eqno(2.4)$$

\n Being a quadratic a functional integral over $\O ({\bf x}, t)$,
 it can be explicitly evaluated in a non-perturbative
manner.

\n The quantum state of the universe is a wavefunctional
$\Psi [ \O, t] $. The probability amplitude
$K [ \O_i; \O_f]$ is the propagator  which gives the quantum state
 $\Psi_f$ at some
time $t_f$ given $\Psi_i$ at time $t_i$ through the relation

$$ \Psi [\O_f] = \int {\cal D} \O  ~K [ \O_i; \O_f]
{}~\Psi_i[\O_i]~.\eqno (2.5)$$

\n The transition amplitude between $\Psi_i$ and $\Psi_f$ is given
 by

$$ \la \Psi_f | \Psi_i \ra = \int\int {\cal D} \O_1  {\cal D} \O_2
{}~\Psi_f^*[\O_2]~ K [ \O_2; \O_1]
 	 ~\Psi_i [ \O_1] ~. \eqno (2.6) $$

\n The `creation' of the universe as a quantum event (like vacuum
 fluctuations,  tunnelling etc.) has been
 explored  by many authors from various points  of view (~see for
example, Tryon 1973; Brout 1980; Zeldovich 1981;
 Atkatz 1982; Vilenkin 1982 and for a review, Kandrup and Mazur 1991~).
 In this framework, the `creation' of the universe has been studied
by evaluating the transition probability from a state  $\Psi_i$ peaked
 around the flat Minkowski spacetime $(\O_i \equiv 1 $
and $ g_\mn(\G_{cl}) = \eta_\mn$) to some state $\Psi_f$ around a
conformally flat spacetime $\Ora \equiv \O ({\bf x} ,t)$.
 It is found that the former is unstable to such fluctuations
(~Atkatz and Pagels 1982; Brout et al 1980; Padmanabhan 1983~) and
 the transition probability between  $\Psi_i $ --- a gaussian wave
 packet around
$\Ora_i$  and a state $\Psi_f$ --- a gaussian packet around
$\Ora_f = \O_f ({\bf x} , t) $ can be written
as

$$ |\la \Psi_f | \Psi_i \ra |^2 = {\cal N } ~ {\rm exp }~
\bigg[ - {1 \over 2 } {\cal W} \bigg]~, \eqno (2.7a)
$$

\n where

$${\cal W} =  {1 \over l_p^2}~\int \int { \nabla \O_f ({\bf x}_1) .
 \nabla \O_f({\bf x}_2 ) \over | {\bf x}_1 - {\bf x}_2|}
{}~ d^3 {\bf x}_1 ~d^3 {\bf x}_2~~, \eqno (2.7 b) $$

\n and $l_p$ is the planck length. The probability of transition is
maximum when ${\cal W } = 0 $ (since $ {\cal W} \geq 0 $) which
occurs for
$\nabla \O_f = 0 $. This implies that $\O_f \equiv \O_o (t) $ which
corresponds to a flat FLRW metric. Hence, we
see that a $\Psi_f $ peaked around the  flat FLRW universe,
 $ \Ora \equiv \oo$, is the most probable outcome of a
causal structure preserving quantum transition of the universe
 from a `ground' state (~peaked around the flat Minkowski
geometry~). The resultant universe being a strictly flat FLRW model
 (i.e., $\oo \equiv 1$, rather than $\oo \approx 1$ as in inflation),
 this approach is free from the
type of criticism of Ellis (1988) against inflation.

%%%%%%%%%%%%%%%%%%%%%%%%%%%%%%section
% 3%%%%%%%%%%%%%%%%%%%%%%%%%%%%%%%%%%%%%%%%%%%%%%%%%%%%%%%%%%%%

\section { 3. The generation of inhomogeneties }

In this section, we  build upon the result reviewed in $\S$2 to
 generate cosmological perturbations.
The exponent ${\cal W}$ of the transition probability  given by
equation (2.7b), can be rewritten in the momentum space as

$$ {\cal W} = {1 \over l_p^2}~\int |Q_k (t)|^2 |{\bf k}|
{ d^3 {\bf k} \over (2 \pi)^3} \eqno (3.1a) $$

\n where

$$\O_f ({\bf x }, t) = l_p^{3 \over 2}~\int Q_{\bf k } (t)
{\rm e}^{i {\bf k }. {\bf x}} { d^3 {\bf k} \over (2 \pi )^3}~.
  \eqno (3.1b) $$

\n Clearly, the homogeneous mode with $ Q_k =0 $  for
 $|{\bf k}| \neq  0$  (i.e., $ \O_f \equiv \O_f (t) $) leads to
 the maximisation
of $|\la \Psi_f | \Psi_i \ra |^2 $. However, it is equally obvious
that in the `next best' situation a
transition to $\O_f$, with weak
inhomogeneties can occur with a slightly reduced probability
 still quite close to unity.

\n Suppose that, in the final state of process of  the quantum
creation, the universe has $\Psi_f$  peaked around

$$ \la \O ({\bf x}, t) \ra_f = \O_o(t) + \e \p \eqno (3.2)$$

\n where $\e $ is a small number. The $\la \O ({\bf x}, t) \ra $
in the above expression, involves a  transition
from the regime
of quantum fluctuations  to classical  perturbations. In our case
 this identification of a classical
field with the expectation value of quantum fluctuations of a
quantum field is very well defined owing
to the fact that the expectation values are calculated between
 coherent quantum states.

\n The probability  of transition ${\cal P }$ to a given $\Ora_f$
can be evaluated using equation $(2.7)$ and $(3.2)$.
 $ {\cal W} $ now reads

$$ {\cal W} = {\e^2 l_p \over 2}~\int |q_{\bf k} (t)|^2 |{\bf k}|
{ d^3 {\bf k } \over (2 \pi)^3} \eqno (3.3a) $$

\n with

$$\p = l_p^{3 \over 2} \int q_{\bf k} ~e^{i {\bf k} . {\bf x}}
 { d^3 {\bf k} \over (2 \pi )^3}~~.  \eqno (3.3b) $$

\n The dependence of ${\cal P}$ on the length scale of the
 inhomogeniety can obtained  from  ($3.3$) by substituting  a form of
$q_{\bf k}$ with a characteristic scale built into it.  We take
 $q_{\bf k}$ to be of the form

 $$ q_{\bf k} = ({4\pi \lambda})^{3 \over 2}~ {\rm exp}~[-{k^2
 \lambda^2 \over  2}]~, \eqno (3.4) $$

\n a gaussian around ${\bf k } = 0$ with a characteristic spread
 $\lambda^{-1}$ in  $k$ ( $ \equiv |{\bf k}|$).

\n The expression for ${\cal W}$ now becomes

$$ {\cal W} = {16 \pi \e^2 l_p \over \lambda} \eqno (3.5) $$

\n and the transition probability ${\cal P }$ given by equation
(2.7a) is

$$ {\cal P} \equiv |\la \Psi_f | \Psi_i \ra |^2 = {\cal N }
 ~{\rm exp}~ [ -{\lambda_o \over \lambda}],
	~~~~~~~~~( \lambda_o = 16\pi\e^2 l_p)~. \eqno(3.6) $$

\n The expression (3.6) shows that the generation of inhomogeniety
at small enough length scales ($\lambda  \ll \lambda_o $)
 is exponentially suppressed.

\n The issue which is addressed next is to estimate the power on
 various scales of inhomogeniety. We calculate the energy content,
 ${\cal E}$, of the final wavepacket $\Psi_f$ in terms of the Fourier
 modes $q_{\bf k} (t) $ of $ \p $
. The expectation value of the energy density  of the wavepacket
$\Psi_f$ is given by

$$ \langle \Psi_f | T_{00} ({\bf x}) |\Psi_i\rangle \equiv
\langle T_{00} ({\bf x})\rangle = \int { \cal D }\O ~ \Psi_f^*[\O]~
 \hat {\cal  H}_\O \Psi_f [\O]~, \eqno (3.7)$$

\n where $\hat {\cal H}_\O$ is the Hamiltonian operator. The above
functional integral can be evaluated explicitly ~(see appendix).

\n The energy $ {\cal E}$ of the wavepacket, obtained by
 integrating $\langle T_{00} ({\bf x})\rangle $
over all space, is of the form

$$ {\cal E} = \int d^3 {\bf x} ~\langle T_{00} ({\bf x})\rangle
 ~~={\e \over 6 l_p^2} \int {d^3 {\bf k} \over (2\pi)^3} k^2
	q_{\bf k} q_{-\bf k}~.  \eqno (3.8)$$

\n We estimate the energy at a scale $\lambda$  by evaluating the
 expectation value $\la {\cal E}^2\ra $
(as outlined in the appendix) and arrive at the result that

$$ \la {\cal E}^2\ra ~ \propto ~~\lambda^{-4}~.\eqno(3.9) $$

\n A measure of power $P(\lambda)$, at a scale $\lambda$ can then
 be taken to be the energy of a
wavepacket with a characteristic scale of inhomogeniety $\lambda$,
 weighted by the relative
probability of transition to such a state. The power spectrum so
 defined, reads

$$ P(\lambda) \propto \lambda^{-4}
{\rm exp}~[- {\lambda_o \over \lambda}] ~.\eqno (3.10) $$

\n The power spectrum has a power law distribution for large
 wavelength modes

$$ P(\lambda)~ \propto~  \lambda^{-4} ~~~~~~~~~~~
(\lambda \gg l_p) \eqno (3.11) $$

\n with a exponential  cut off at small values of $ \lambda $.
The peak occurs at a wavelength $\lambda_{\rm peak} $,

$$\lambda_{\rm peak} = {1 \over 4} \lambda_o
{}~= 4\pi \e^2 l_p~. \eqno (3.12) $$

\n The above results show an interplay between the wavelength
$\lambda_{\rm peak}$ at which inhomogeneity is generated and the
amplitude $\e$.

\n We  investigate the nature of these perturbations by looking at
 the gauge invariant potential
$\Phi_H$, introduced by Bardeen (1980), in a universe filled with
 two component (radiation and dust) hydrodynamic matter.
The conformal fluctuations $\p$ would not cause any deviation of
 the perturbed spacetime from conformal flatness,
 implying that the Weyl curvature tensor $ C_{\mn \lambda\d} $ is
 identically zero. The Bardeen potential as a
 geometrical quantity is proportional to the square of the Weyl
 curvature,

$$ \Phi_H ~ \propto ~C_{\mn\lambda\d} C^{\mn\lambda\d}, \eqno (3.13) $$

\n and relates to the matter pertubations
(~Starobinsky and Sahni 1984~) as

$$\Phi_H  \propto ~ \d \rho_{\rm tot.} ~~(=~ \d \rho_{\rm radn.}
+ \d \rho_{\rm dust}) ~.\eqno (3.14) $$

\n The above equations (3.13) and (3.14), coupled with the fact
 that the Weyl tensor $ C_{\mn \lambda\d} \equiv 0 $
for conformal fluctuations leads to the conclusion that

$$ \d \rho_{\rm tot.} = \sum_{\rm i} \d \rho_{\rm i} = 0, \eqno (3.15) $$

\n where we have stated  our result as generalised to
multi-component hydrodynamic matter. From the equation (3.15),
the seed perturbations are recognised to isocurvature perturbations.

\n Isocurvature perturbations are a firm prediction of our scenario.
 The role of isocurvature perturbations in both the
baryon dominated (~Peebles 1987a, 1987b; Efstathiou and Rees  1988~)
 and cold dark matter (CDM) models (~ Efstathiou and Bond 1987;
 Bardeen et al
1987; Starobinsky and Sahni 1984~) of structure formation have
 been discussed in the literature. These perturbations are known to
generate more power on large scales in CDM models which violates
CMBR bounds, but can be exploited in Baryon dominated models
because of characteristic
features appearing in their final spectrum at astrophysically
large scales.

%%%%%%%%%%%%%%%%%%%%%%%%%%%%%%%%%%%%%%%%%%%section 4
% %%%%%%%%%%%%%%%%%%%%%%%%%%%%%%%%%%%%%%%%%%%%%%%%%%%%%%

\section{4. Translation to mass perturbations }

\n The early universe is expected to be hot i.e., matter is in
thermal equilibrium at a very high temperature.
At high temperatures, all particles would be relativistic hence
 one can consider the matter Lagrangian to be invariant
under conformal transformations of the metric. However, as the
 universe cools down, some component of matter would break
its conformal invariance. The simplest picture would be to have
some particle become non-relativistic at some epoch $t^*$
(corresponding to some mass scale) or, alternatively, a phase
transition may cause some massless boson to acquire
mass. This component of matter  would then {\it see} the conformal
 fluctuations and chart out
a corresponding density perturbation. A simple analysis shows the
 relation between the ${\delta M / M}$ and
the conformal fluctuations $\p$. In line with the conventional
 treatment of cosmological perturbations (~Landau 1958;
 Mukhanov et al 1991~), we compute
the perturbations $\d M $ in some physical quantity  $M $
 (say mass) as the the difference between the value $M $
 calculated in the physical 3-hypersurface $\Sigma$ with
conformal factor $\O ({\bf x }, t) $,  and the
 homogeneous background value $\overline M $  calculated in
the background 3- hypersurface~$\overline \Sigma $.

\n Consider a comoving 3-volume $\D V_c$. The mass $M (\D V_c) $
 contained in the corresponding physical
volume  is given by

$$ M (\D V_c)   = \O^3 ({\bf x }, t)  \rho~\D V_c =
{}~[ \oo + \e \p ]^3 \rho~\D V_c, \eqno (4.1) $$

\n and the mass ${\overline M }(\D V_c) $ contained in the
background volume is

$$ {\overline M }(\D V_c) = \O_o^3(t)  \rho \D V_c ~.\eqno (4.2) $$

\n As noted in $\S$3, the perturbations are of the isocurvature
 kind,  and hence the $\d \rho$ term that one would expect in equation
($4.1$) is absent. The mass fluctuation at a point,
 ${ \delta M / M} (  x , t)$ can be expressed
 (upto linear order in $\e$) as

$$ { \delta M \over M} ( {\bf x }, t) = \eta
{}~\bigg[{ M (\D V_c) - {\overline M }(\D V_c)  \over
{\overline M }(\D V_c)}\bigg]
= 3\e \eta {\p \over \oo} ~,\eqno (4.3) $$

\n where the constant $\eta ~( \leq 1) $ has  been introduced
to  account for the fact that only a fraction of the
matter would respond to the conformal fluctuations. The mass
 fluctuations $ { \delta M / M} $ are directly
proportional to the conformal fluctuations $\p$. In the
 Fourier space, the mass fluctuation at a  scale $k$
(i.e., the power spectrum of ${ \delta M / M} $) would be directly
 related to the  power spectrum of the conformal
perturbations obtained in $\S$3. The power spectrum of the mass
 fluctuations, using ($3.11$) can be written as

$$ \bigl| {\delta M \over M}\bigr|^2 ( {\bf k} , t) = P(k^{-1})
 ~f (t) \propto k^4 {\rm e}^{- \lambda_o k } f (t)~, \eqno (4.4) $$

\n where $f (t)$ incorporates the time dependence  of $\p$ which
 would be taken up in the next section. The framework within which
 we are working does not require us to invoke an inflationary
 stage during the
evolution of the universe; hence, the astrophysically relevant
 scales are the modes with extremely
large wavelengths (~$k^{-1}  \sim 10^{28} l_p $). Therefore,
 the power spectrum of fluctuations
at the astrophysically relevant scales is given by

$$\bigl| {\delta M \over M}\bigr|^2_k \propto k^4
{}~~~~~~~~(k l_p \ll 1). \eqno (4.5) $$

\n This is the  Harrison-Zeldovich spectrum (~ Harrison 1970;
 Zeldovich 1972~).  To state the above more clearly,
 we use

$$ \bigl| { \delta M \over M}\bigr|^2_k
\approx k^3 |\d_k|^2 \eqno (4.6a) $$

\n where

$$ \d_k = \int d^3 {\bf x} ~~e^{i{\bf k}.{\bf x}}
{}~{\d \rho \over \rho}({\bf x }, t). \eqno (4.6b) $$

\n Using ($4.5$) and ($4.6$) we get

$$ |\d_k|^2 \propto k ~~~~~~~~(k \ll 1), \eqno(4.7) $$

\n which is the Harrison-Zeldovich spectrum in the
 more familiar form.

%\vfill\eject

%%%%%%%%%%%%%%%%%%%%%%%%%%%%%%%%%%%%%%%%%%% section 5
% %%%%%%%%%%%%%%%%%%%%%%%%%%%%%%%%%%%%%%%%%%%%%

\section{5. Evolution of fluctuations}

We now address the question of evolution of the conformal
 fluctuations  generated at the time of the `creation'
of the universe till the time they decay away into seed
 perturbations in the non-conformal component of the matter
in the universe. After the quantum transition has occured,
the conformal factor can be treated as a classical field
with an action $J$ as given in equation ($2.2$) evolving in
a background geometry. Our analysis would be only limited to
 the initial evolution of conformal fluctuations  which later
 decay away transfering their energy to density
perturbations (as outlined in $\S$3) and we assume that these
follow the well known evolution equations (Peebles~1980).

\n The action governing the conformal fluctuations
 $\O ( {\bf x}, t) $ around a fiducial metric
$\tilde g_\mn $ has the form

$$ J = {1 \over 16 \pi } \int [6 \O^\mu \O_\mu - \tilde R \O^2]
 \sqrt {- \tilde g} ~ d^4x + J_m \eqno (5.1) $$

\n The equation of motion of the conformal fluctuation
$\O({ \bf x}, t )$ is

$$ \tilde {\dal }~ \O + {1 \over 6} \tilde { R} \O
= {\delta {\cal L}_m \over \delta \O}~~~; \eqno (5.2)$$

\n akin to that of a conformally coupled scalar field with
 a potential $V(\O)$ (~Narlikar and Padmanabhan 1983~).

\n In particular, for  $\tilde g_\mn \equiv \eta_\mn $,
equation ($5.2$) reduces to

$$ \pl_\mu \pl^\mu \O({\bf x }, t) =
{\delta {\cal L}_m \over \delta \O}
 \equiv V^\prime(\O)~. \eqno (5.3) $$

\n At this stage, we go ahead with the idea  that the
 matter Lagragian is invariant under
conformal transformations (i.e., set $V^\prime (\O) \equiv 0$).
 Substituting the expression ($3.2$) into equation ($5.3$),
 we get at the zeroth order

$$ {d^2 \over dt^2}\oo = 0 ,~~~\Rightarrow
{}~~~\oo \propto t\eqno (5.4a)$$

\n and to the first order in $\e$,

$$ \dal ~\p = 0 ~.\eqno (5.4b) $$

\n The equation ($5.4b$) admits plane travelling wave solution
 for $\p$.

\n However, in a more realistic treatment one would have a
non-trivial form for $V(\O)$. We will bypass
the question of determining the exact form of $V(\O)$ and
assume a form

$$ V(\O) = m^2 \O^2 \eqno(5.5) $$

\n where $m$ is the mass scale introduced to break the conformal
 invariance of ${\cal L}_m$. Using the form
of $V(\O)$ given in ($5.5$) in the equation ($5.3$), we obtain
at the zeroth order

$$ {d^2 \over dt^2} \oo  = m^2 \oo \eqno(5.6a) $$

\n and at  ${\cal O}( \e)$

$$  \dal ~\p = m^2 \p. \eqno (5.6b) $$

\n The form of ($5.6b$) suggests that the Fourier components
of $\p$ would  obey the evolution equation

$$ \phi_{\bf k} (t)  \propto
\exp~ [\pm (m^2 - |{\bf k}|^2)^{1 \over 2}~t]~ .\eqno (5.7) $$

\n  The inhomogeneous modes $ \phi_k(t)$, have a growing solution
for $k \leq m $ (and also a decaying
solution). This implies that in the case of $V(\O)$ of the form ($5.5$), the
mass scale $m$ introduces
 a lower cut off value  for the wavelength above which one
finds growing modes. This further strengthens
our argument that  inhomogeneties would exist only at the
large wavelength modes which in our framework
are the astrophysically relevant scales.

%%%%%%%%%%%%%%%%%%%%%%%%%%%%%%%% section 6
% %%%%%%%%%%%%%%%%%%%%%%%%%%%%%%%%%%%%%%%%%%%%%%%%%%%%%%%%%%%%%%

\section{6. Alternative prescription for evolution }

In the preceding section, the evolution of the conformal
fluctuations was governed by equation (5.2) where the
potential $V (\O) $ had to be fixed from some independent
 physical considerations regarding the coupling of matter to
the conformal degree of freedom of gravity. The analysis of
 Narlikar and Padmanabhan (1983) outlined in $\S$2, indicated
that the conformal factor would be homogeneous at the leading
order. However, the exact form of $\oo$ is not uniquely determined.
 This opens up the possibility of an alternative  prescription
where one could fix the form of
$\oo$ first and then consider the evolution of $\p$ through
equation ($5.3$) using the corresponding form for $V(\O)$.
In this section, we outline one possible physical consideration
through which $\oo$  could be obtained.

\n We rewrite the background metric in comoving coordinates as

$$ ds^2 = \O^2 (t)~ (dt^2 -d{\bf x}^2) = d\T^2 - S^2(\T)~
 d{\bf x}^2 \eqno (6.1) $$

\n and assume that the energy density of the universe in the early
 epoch would be given by
some form of an uncertainity relation. The Heisenberg's uncertainity
 relation between
energy and time, ( $\D E .~\D \T = 1 $ ), in the context of a
 quantum universe  would involve the total
energy  and the time elapsed since its creation. The total energy,
 $E $ in a physical volume $V_{\rm phy.}$ can be expressed as

$$ E = T_{00} ~V_{\rm phy.} = T_{00} ~S^3(\T) V_c \eqno (6.2a) $$

\n where $V_c$ is a constant. At early epoch
(around the planck era)

$$\D \T \approx \T,  \eqno (6.2b)$$

\n which in conjunction with the uncertainity principle and
 equation (6.2) would lead to the following  relation

$$ T_{00} = {K \over \T S^3(\T) } ~, ~~~~~~
(K \equiv constant)~. \eqno (6.3) $$

\n  Now we try to bridge the gap between the quantum universe
 created and the classical universe that follows later, by
postulating that the classical homogeneous FLRW spacetime that
emerges is  dictated by the energy density given
by equation (6.3).  For a flat FLRW metric ($6.1$), equation
($6.3$) takes the form

$$ {1 \over S^2(\T)} \bigg( {dS(\T) \over d\T} \bigg)^2
= {K^3 \over S^3(\T) \T}~; \eqno(6.4a) $$

\n which determines $S(\T) $ (assuming $S(\T) =0 $ at $\T =0$ )
to be

$$ S(\T) = K^{1 \over 3} \T^{1 \over 3}. \eqno (6.4b) $$

\n The above corresponds to the case of a universe filled with a
 hydrodynamic fluid with a stiff equation
of state ($p = \rho$). The corresponding conformal factor
$\oo$ reads

$$ \oo = 2 {\sqrt 3} K^{1 \over 2} t^{1 \over 2} ~. \eqno (6.4) $$

\n This would also change the evolution equation for $\p$ since
the form of $T_{00}$ given by $(6.1)$
would imply a particular form of $V(\O)$ in $(5.3)$. The above
 consideration is more in the vein of illustration
and one can generalise it to $\O_o$ of the form $\oo \propto
 t^\gamma$ to
cover a broader range of possibilities
(e.g. $ \gamma = -1 ~ \Rightarrow $ De Sitter spacetime).

%%%%%%%%%%%%%%%%%%%%%%%%%%%%%%% section 7
% %%%%%%%%%%%%%%%%%%%%%%%%%%%%%%%%%%%%%%%

\vfil\eject
\section { 7. Discussion }

\n We have extended the concept of quantum conformal cosmology
 originally put forward by Narlikar and Padmanabhan
 (1983) to resolve the problems of the singularity, horizon and
flatness problems in classical cosmology, to study the
origin of density perturbations in our universe.

\n The scenario outlined harnesses the quantum fluctuations in
 the gravitational sector (~e.g., see Halliwell 1985~)
in contrast to the inflationary ones which use quantum fluctuations
in the matter sector. Futhermore, in our case the perturbations
are not generated  through microphysical processes, in fact the
 universe is `born',
with a high probability, as a flat FLRW spacetime with weak
 inhomogeneities on large scales.
This approach therefore circumvents the need for inflation
 wherein causally produced small scale quantum fluctuations
are stretched to
exponentially large wavelengths which are then argued to be
 mimicing a classical, weakly inhomogeneous field
(~Vilenkin 1983; Brandenberger 1990~). Furthermore, the
identification of a classical field with the expectation values
of the quantised (conformal) field is well defined in our scenario
 since we are dealing with coherent states.

\n  The power spectrum of the inhomogeneties, $\p$, in the conformal
 factor is  directly related to the
mass fluctuations ${\d M / M } $ in the conformally non-invariant
 component of matter. This leads
to a scale invariant spectrum ({\it Harrison-Zeldovich spectrum })
for ${\d M / M } $ for the
large wavelength modes which in our case are the astrophysically
relevant scales.  The conformal fluctuations
are recognised to be  of  the isocurvature kind and  here we have
studied their evolution
 for conformal matter as well as for the case of quadratic
coupling of the conformal sector
of gravity to the  matter Lagrangian.

\n At this  stage, we would like to clarify that the often
 quoted unphysical nature of the  conformal
sector of gravity stems from considerations of a Euclideanised
 quantum gravity. The fact that, akin to theories
with guage freedom, the kinetic term in the  conformal sector
 (in euclideanised gravity) appears with a negative sign has
 prompted attempts to factor out the conformal factor and treat
it like  a non-dynamical (unphysical) degree of
freedom (for e.g., see Mazur and Mottola 1990~).
In our  case, we do not use any form of Euclideanisation  and
besides, it is  easy to see that the
 conformal factor is as physical as the scale factor of the
3-hypersurfaces in FLRW spacetime (see equation (6.1)~)
 since the two are related by a gauge transformation (the choice
of the lapse function in a 3+1 York decomposition).

\n  Our discussion in this paper illustrates that quantum conformal
cosmology, albeit arising out of a simplistic model of quantum gravity,
 can
address all the  long standing problems in big bang cosmology.
We would conclude that
though inflation is an attractive concept, it is not indispensable
and one should keep an open mind for other
alternative scenarios for solving the outstanding problems of
classical big  bang cosmology.

%%%%%%%%%%%%%%%%%%%%%%%%%%%%%%%%%%%%% acknowledgement
% %%%%%%%%%%%%%%%%%%%%%%%%%%%%%%%%%%%%%%%%%%%%%%%%%%%%%

\section {Acknowledgement}

It is a pleasure to thank Prof. Jayant Narlikar for his guidance
and encouragement.
The author greatly benefitted from very fruitful discussions with
  T. Padmanabhan and also thanks K. Subramanian,
Varun Sahni and Abhijit Kshirsagar for their useful suggestions
 during the course of this work.
This work was financially supported by the Council of Scientific
and Industrial Research, India under its SRF scheme.

\bigskip
\bigskip

\vfill\eject

%%%%%%%%%%%%%%%%%%%%%%%%%%%%% Appendix
% %%%%%%%%%%%%%%%%%%%%%%%%%%%%%%%%%%%%%%%%%%%%%%%%%%%%%%%%%%

\section {Appendix :  Energy in a wavepacket }

Consider a state $\Psi_f [\O]$, a functional of the conformal factor,
 to be strongly peaked around
$\la \O\ra_f = \oo + \e \p $.  In the Fourier domain, this state is
 represented by a functional
of the Fourier coefficients $Q_{\bf k}$ of $\O ({\bf x},t)$ as

$$\tilde \Psi_f [Q_{\bf k}]  = {\cal N} \exp \bigg[ - {3 \over 8 \pi}
\int |  Q_{\bf k} -  q_{\bf k}
	|^2 ~{ d^3 {\bf  k} \over (2\pi)^3}\bigg] \eqno (A.1)  $$

\n  which is peaked around $q_{\bf k}$, the Fourier coefficient of $\p$.

\n The expression for $\la T_{00} ({\bf x})\ra $ given in equation
 ($3.7$), rewritten in the Fourier domain reads

$$ \langle T_{00} ({\bf x})\rangle =	\int { \cal D }\O ~
 \Psi_f^*[\O]~ \hat {\cal  H}_\O
\Psi_f [\O]= \int { \cal D }Q_{\bf k} ~ \tilde \Psi_f^*[Q_{\bf k}]~
 \hat {\cal  H}_{Q_{\bf k}} \tilde \Psi_f [Q_{\bf k}] ~, \eqno (A.2)$$

\n where the Hamiltonian operator,

$$ \hat {\cal  H}_\O  = - { l_p^2 \over 2} { \d^2 \over \d \O^2}
 + {1 \over 6 ~l_p^2} \O_i \O^i + {1 \over 12~l_p^2} R \O^2.
	\eqno (A.3)  $$

\n  Since we are dealing with spacetimes which are close to $ R =0 $, the third
term will be ignored in the following calculation.

 In the Fourier domain

$$ \hat {\cal  H}_{Q_{\bf k}}	= \int\int {d^3 {\bf  k}_1 \over
 (2\pi)^3}~{d^3 {\bf  k}_2 \over (2\pi)^3}~
	e^{i ({\bf k}_1 + {\bf k}_2). {\bf x}} ~\bigg[ - { l_p^2 \over 2} { \d^2 \over
\d Q_{{\bf k}_1} \d Q_{{\bf k}_2}} + {1 \over 6 ~l_p^2}{\bf k}_1.{\bf k}_2~
Q_{{\bf k}_1} Q_{{\bf k}_2} \bigg]. \eqno (A.4) $$

\n  The expectation value of the energy density can be evaluated
by substituting equations ($A.1$) and  ($A.4$) into equation
 ($A.2$). The  steps in the calculation involve standard functional
differentiations and the following  result of functional
integration (Freidrich  1976),

$$\eqalign { \int {\cal D} \t (s) \bigg( \prod_s  \sqrt {{\mu (ds)
\over 2\pi }} ~{\rm e}^{-{1 \over 2} \t^2 (s) \mu (ds)} &\int \int
  \mu (ds_1) \mu (ds_2)~ B (s_1, s_2 ) ~\t (s_1) \t (s_2) \cr
   &= \int \mu (ds)~ B (s ,s)~, \cr}  \eqno(A.5a) $$

\n where $ \mu (ds) $ is  a measure defined such that

$$ \int_{- \infty}^{\infty} \sqrt {{\mu (ds) \over 2\pi }}
{}~{\rm e}^{-{1 \over 2} \t^2 (s)~ \mu (ds)} = 1~. \eqno(A.5b)$$

\n In our case, $\tilde \Psi_f [Q_{\bf k}] $ being a normalised
gaussian wavepacket, equation ($A.5$) leads
to a useful result

$$ \int {\cal D} f_{\bf k} ~\tilde \Psi_f [f_{\bf k}] ~~\int\int
{d^3 {\bf  k}_1 \over (2\pi)^3}~{d^3 {\bf  k}_2 \over (2\pi)^3}
{}~ B(k_1, k_2) ~ f_{{\bf k}_1} f_{{\bf k}_2} = \int {d^3 {\bf  k}
\over (2\pi)^3}~B ({\bf k},{\bf k})~.\eqno (A.6) $$

\n  where $f_{\bf k} = |  Q_{\bf k} -  q_{\bf k}| $. The final
 expression for $\langle T_{00} ({\bf x})\rangle$ appears as a
sum of integrals in the Fourier domain as

$$ \langle T_{00} ({\bf x})\rangle =  {\e^2 \over 6~l_p^2}\int
 {d^3 {\bf  k}_1 \over (2\pi)^3}~{d^3 {\bf  k}_2 \over (2\pi)^3}~
	{\rm e}^{i ({\bf k}_1 + {\bf k}_2). {\bf x}} ~{\bf k}_1.
{\bf k}_2~ q_{{\bf k}_1} q_{{\bf k}_2}
+ {\rm (other ~terms) }~, \eqno (A.7)$$

\n  where only the  term that contributes to the energy
${\cal E}$ is written out explicitly. The energy

$$ {\cal E} = \int  d^3 {\bf  x } ~\langle T_{00} ({\bf x})\rangle
 =  {\e^2 \over 6~l_p^2} \int {d^3 {\bf  k} \over (2\pi)^3}
	 k^2 q_{\bf k}~ q_{- {\bf k}}~, \eqno (A.8) $$

\n is obtained using equation ($3.8$), by integrating
 $ \langle T_{00} ({\bf x})\rangle$ over all space.

\n We arrive at the result that, corresponding to every
 $\O_f ({\bf x}, t)$, there exists an energy functional
 ${\cal E} [\O_f]$ or
equivalently ${\cal E} [ q_{\bf k}]$. The equation ($A.8$)
 can be expressed  formally as

$$ {\cal E} = \sum_{\bf k} {\e^2 \over 8\pi}
k^2 q^2_{\bf k} ~~\equiv \sum_{\bf k} {\cal E}_{\bf k}~. \eqno (A.9)$$

\n Clearly, ${\cal E}$ is a Gaussian random variable with a
 probability distribution proportional to that of $\O_f ({\bf x}, t)$

$$ {\cal P} [{\cal E}] \propto ~{\cal P} [ q_{\bf k}]
 = \prod_{\bf k} ~~\exp~\bigg[ - 16\pi^2 {{\cal E}_{\bf k}
	 \over |{\bf k}|}\bigg]~,\eqno(A.10) $$

\n where we have used equations ($ 3.3$) and ($A.9$).

\n The mean square fluctuation in energy in $\p$ would be

$$ \la {\cal E}^2 \ra ~= \int  {d^3 {\bf  k} \over (2\pi)^3}
\bigg( { k \over 16\pi^2}\bigg)~\propto \lambda^{-4} ~,\eqno(A.11)$$

\n $\lambda$ being a cut off length scale for  the above integral.

\vfil\eject

%%%%%%%%%%%%%%%%%%%%%%%%%%%%% References
% %%%%%%%%%%%%%%%%%%%%%%%%%%%%%%%%%%%%%%%%%

\section { ~~~~References }

\I {} Atkatz, D. and Pagels, H. 1982, {\it  Phys. Rev.}  {\bf D25}, 2065.

\s

\I{} Bardeen, J. M. 1980, {\it  Phys. Rev.} {\bf D22}, 1882.

\s

\I{}  Bardeen, J. M., Steinhardt, P. J., and Turner, M. S. 1983,
 {\it   Phys. Rev.} {\bf D28},
 679.

\s

\I{} Bardeen, J. M., Bond, J. R., and Efstathiou, G. 1987,
 {\it  Astrophys. J.} {\bf 321}, 28.

\s

\I{}  Brandenberger, R. H. 1985, {\it  Rev. Mod. Phys.} {\bf 57}, 1.

\s

\I{} Brandenberger, R. H., Laflamme, R., and Mijic, M. 1990,
{\it  Mod. Phys. Lett.} {\bf A5}, 2311.

\s

\I{} Brout, R., Englert, F., and  Gunzig, E., 1978,
{\it  Ann. Phys. (N.Y.)} {\bf 115}, 78.

%\s

\I{} Brout, R., Englert, F., Frere, J. M., Gunzig,
E., Naradone, P., and Truffin, C. 1980,
	{\it  Nucl. Phys. } {\bf B170}, 228.

\s

\I{} Dicke, R. H. and Peebles, P. J. E. 1979,
 in {\it General Relativity} : {\it An Einstein Centenary
	 Survey} ~ ed. S. W. Hawking and W. Israel (Cambridge University Press) p.
504.

\s

\I{} Efstathiou, G. and Bond, J. R. 1987,
 {\it   Mon. Not. R. astron. Soc.} {\bf 227}, 33p.

\s

\I{} Efstathiou, G. and Rees, M. J. 1988,
 {\it  Mon. Not. R. astron. Soc.} {\bf 230}, 5p.

\s

\I{} Ellis, G. F. R.  and Stoeger, W. 1988,
{\it  Class. Quantum Grav.} {\bf 5}, 207.

\s

\I{} Ellis, G. F. R. 1988, {\it  Class. Quantum Grav.}
{\bf 5},  891.

\s

\I{} Feynman, R. P. and Hibbs, A. R. 1965,
 {\it Quantum Mechanics and Path Integrals},
	(Mc.Graw-Hill, New York).

\s

\I{} Freidrichs, K. O., Shapiro, H. N. et al 1976,
{\it Courant Institute Lec. Notes} :
	{\it Integration of Functionals} (Courant Institute of Math. Sciences, New
York University).

\s

\I{} Guth, A. H. 1981, {\it  Phys. Rev.} {\bf D23}, 347.

\s

\I{} Guth, A. H. and Pi, S. Y. 1982, {\it  Phys. Rev. Lett.}
 {\bf 49}, 1110.

\s

\I{} Halliwell, J. J. and Hawking, S. W. 1985,
{\it  Phys. Rev.} {\bf D31}, 1777.

\s

\I{} Harrison, E. 1970, {\it  Phys. Rev.}
{\bf D1}, 2726.

\s

\I{} Hawking, S. W. 1982, {\it  Phys. Lett.} {\bf 115B}, 295.

\s

\I{} Isenberg, J. and Wheeler, J. A. 1979,
 in {\it Relativity, Quanta and Cosmology}, ed.~M. Pantaleo
	 and  F. de Finis, (Johnson, New York 1979) p. 267.

\s

\I{} Kandrup, H. E. and Mazur, P. O. 1991,
 {\it  Int. Jour. Mod. Phys.} {\bf A6}, 4041.

\s

\I{} Landau, I. and Lifshitz, E., {\it Theoretical Physics Vol. II}
: {\it Classical Fields},
	(Pergamon, London 1958).

\s

\I{} Linde A. D. 1982, {\it Non singular regenerating
 inflationary universe}, Cambridge University
	{\it preprint}.

\s

\I{} Madsen, M. S. and Ellis, G. F. R.  1988,
{\it   Mon. Not. R. astro. Soc.} {\bf 234}, 67.

\s

\I{} Mazur, P. O. and Mottola, E. 1990, {\it  Nucl. Phys.}
 {\bf B341}, 187.

\s

\I{} Misner, C. W. 1969, {\it  Phys. Rev. Lett.}
  {\bf 22}, 1071.

\s

\I{} Mukhanov, V., Feldman H., and Brandenberger,
R. H. 1991, {\it  Phys. Rep.}, {\it in press}.

\s

\I{} Narlikar, J. V. 1979, {\it  Gen. Rel. Grav.} {\bf 10}, 883.

\s

\I{} Narlikar, J. V. 1981, {\it  Found. Phys.} {\bf 11}, 473.

\s

\I{} Narlikar, J. V. and Padmanabhan, T. 1983, {\it Ann. Phys.}
 {\bf 150}, 289.

\s

\I{} Narlikar, J. V. and Padmanabhan, T. 1991,
{\it  Ann. Rev. Astr. Ap.} {\bf 29}, 325.

\s

\I{} Padmanabhan, T. and Narlikar, J. V. 1982,
{\it  Nature } {\bf 295}, 677.

\s

\I{} Padmanabhan, T. 1982, {\it  Phys. Rev.} {\bf D26}, 2162.

\s

\I{} Padmanabhan, T. 1983, {\it  Phys. Lett.} {\bf 93A}, 116.

\s

\I{} Padmanabhan, T. and Seshadri, T. R. 1988,
{\it  Class. Quantum Grav.} {\bf 5}, 221.

\s

\I{} Padmanabhan, T. 1989, {\it  Int. Jour. Mod. Phys.} {\bf A4}, 4735.

\s

\I{} Peebles, P. J. E. 1980,
 {\it The Large-Scale Structure of the Universe},
	(Princeton University Press, New Jersey)

\s

\I{} Peebles, P. J. E. 1987a, {\it Asrophys. J.} {\bf 315}, L73.

\s

\I{} Peebles, P. J. E. 1987b, {\it   Nature} {\bf 327}, 210.

\s

\I{} Starobinsky, A. A. 1980, {\it  Phys. Lett.} {\bf 91B}, 99.

\s

\I{} Starobinsky, A. A. 1982, {\it  Phys. Lett.} {\bf 117B}, 175.

\s

\I{} Starobinsky, A. A. and Sahni, V. in
 {\it Modern Theoretical and Experimental Problems of General
	Relativity} (Moscow, 1984), p. 77; English translation
appeared in 1986 as New Castle  University {\it preprint} $NCL-TP12$.

\s

\I{} Tryon, E. P. 1973, {\it  Nature } {\bf 246}, 396.

\s

\I{} Vilenkin, A. 1982, {\it  Phys. Lett.} {\bf 117B}, 25.

\s

\I{} Vilenkin, A. 1983, {\it  Nucl. Phys.} {\bf B226}, 527.

\s

\I{} York Jr., J. W. 1972, {\it  Phys. Rev. Lett.} {\bf 28}, 1082.

\s

\I{} Zeldovich, Ya. B. 1972, {\it  Mon. Not. R. astr. Soc.}
{\bf 160}, 1.

\s

\I{} Zeldovich, Ya. B. 1981, {\it  Pis'ma Astron. Zh.} {\bf 7}, 579.

\vfill\eject

\end